\documentclass[showpacs,10pt,twocolumn,prb]{revtex4}
\usepackage{amsmath}
\usepackage{amssymb}
\usepackage{graphics}
\usepackage{epsfig}
\usepackage{color}

\begin{document}

\title{Single crystal growth and superconductivity of Ca(Fe$_{1-x}$Co$_{x}$)$%
_{2}$As$_{2}$}
\author{Rongwei Hu$^{\ast }$, Sheng Ran, Sergey L. Bud'ko, Warren E.
Straszheim and Paul C. Canfield}
\affiliation{Ames Laboratory, U.S. DOE and Department of Physics and Astronomy, Iowa
State University, Ames, IA 50011, USA}
\date{\today }

\begin{abstract}
We report the single crystal growth of Ca(Fe$_{1-x}$Co$_{x}$)$_{2}$As$_{2}$ (%
$0\leq x\leq 0.082$) from Sn flux. The temperature-composition phase diagram
is mapped out based on the magnetic susceptibility and electrical transport
measurements. Phase diagram of Ca(Fe$_{1-x}$Co$_{x}$)$_{2}$As$_{2}$ is
qualitatively different from those of Sr and Ba, it could be due to both the
charge doping and structural tuning effects associated with Co substitution.
\end{abstract}

\maketitle

The AEFe$_{2}$As$_{2}$ (AE = Ca, Sr, Ba) of the 122 family is the most
extensively studied materials among the various iron arsenic
superconductors, since they possess the characteristic tetrahedrally
coordinated square planar Fe sublattice, giving rise to lattice instability,
antiferromagnetism (AFM) and superconductivity (SC) by chemical
substitution, and are readily obtained in large single crystalline form.\cite%
{Rotter}$^{-}$\cite{Izyumov} CaFe$_{2}$As$_{2}$ is similar to the other two
members, it undergoes a phase transition from a high temperature, tetragonal
phase to a low temperature, orthorhombic/antiferromagnetic phase below 170 K.%
\cite{Ni2} Superconductivity can be induced in CaFe$_{2}$As$_{2}$ by
substituting Fe with Co\cite{Kumar} or Rh\cite{Qi} and As with P\cite%
{Kasahara} and by application of non-hydrostatic pressure\cite{Canfield1}$%
^{,}$\cite{Milton}, as the tetragonal-orthorhombic/AFM transition is
suppressed, strongly suggesting the connection between the AFM fluctuations
and SC. However, the physical properties of the single crystals of CaFe$_{2}$%
As$_{2}$ are remarkably dependent on the crystal growth procedure. It has
been shown that crystals quenched from high temperature using FeAs flux
exhibit a transition from a high temperature, tetragonal to a low
temperature, non-magnetic, collapsed tetragonal phase below 100 K in
contrast to the behavior of CaFe$_{2}$As$_{2}$ grown from Sn flux.\cite{Ran}
For Co doping, the as grown, single crystals grown from FeAs-CoAs self-flux,
decanted at 1000 $%
{{}^\circ}%
C$, do not show any SC as opposed to the corresponding ones grown from Sn
flux.\cite{Ran2} Moreover, there is a competing phase of CaFe$_{4}$As$_{3}$
growing concomitantly with CaFe$_{2}$As$_{2}$ from Sn flux.\cite{Ni2}
Therefore details in the crystal growth and effects of Co doping in CaFe$%
_{2} $As$_{2}$ need to be clarified. In this work, we performed a study of
the single crystal growth of Ca(Fe$_{1-x}$Co$_{x}$)$_{2}$As$_{2}$ out of Sn
flux and show the dependence of the magnetic susceptibility and resistivity
on Co doping.

Single crystals of Ca(Fe$_{1-x}$Co$_{x}$)$_{2}$As$_{2}$ were grown from Sn
flux in two steps. In order to obtain homogeneous Co substitution for Fe,
polycrystalline Ca(Fe$_{1-x}$Co$_{x}$)$_{2}$As$_{2}$ were prepared first by
heating stoichiometric mixtures of Ca, FeAs and CoAs at 900 $%
{{}^\circ}%
C$ for 24 hours. The polycrystalline sample was ground and pelletized for a
second time sintering at 900 $%
{{}^\circ}%
C$. Then polycrystalline Ca(Fe$_{1-x}$Co$_{x}$)$_{2}$As$_{2}$ and Sn with a
ratio of 1 : 30 were placed in an alumina crucible and sealed in amorphous
silica tubes. The sealed ampoule was heated to 1100 $%
{{}^\circ}%
C$ and slowly cooled to 600 $%
{{}^\circ}%
C$ after which the Sn flux was decanted.\cite{Fisk} This procedure is
similar to the one in Ref. 16. Early work on crystal growth of CaFe$_{2}$As$%
_{2}$ using Sn flux has identified a needle-shaped orthorhombic phase, CaFe$%
_{4}$As$_{3}$, growing together with CaFe$_{2}$As$_{2}$ out of Sn flux.\cite%
{Ni2}\ Our crystal growth showed that there was a significant amount of CaFe$%
_{4}$As$_{3}$ phase by following the above procedure. For Co doping, the
formation of the competing phase may change the composition of the liquid
solution, thus it causes complex dependence of the doping concentration of
the resulted single crystals on growth conditions. An excess of Ca was added
to the polycrystalline and Sn mixture and an optimal Ca$_{1.5}$(Fe$_{1-x}$Co$%
_{x}$)$_{2}$As$_{2}$ was found to be effective for eliminating the CaFe$_{4}$%
As$_{3}$ phase. Moreover, there is a solubility problem of Ca(Fe$_{1-x}$Co$%
_{x}$)$_{2}$As$_{2}$ in Sn.\ For the ratio of 1 : 30, in addition to Ca(Fe$%
_{1-x}$Co$_{x}$)$_{2}$As$_{2}$ single crystals, there was some undissolved
polycrystalline powder after decanting. By changing the ratio to 1 : 45, we
were able to completely dissolve the starting polycrystal. The as-grown
single crystals were thin plate-like with typical dimension $4\times 4\times
0.2$ mm$^{3}$.

Crystals were characterized by powder x-ray diffraction using a Rigaku
Miniflex X-ray diffractometer. The actual chemical composition was
determined using wavelength dispersive x-ray spectroscopy (WDS) in a JEOL
JXA-8200 electron microscope, by averaging ten spots on the crystal surface.
Magnetic susceptibility was measured in a Quantum Design MPMS, SQUID
magnetometer. In plane AC resistivity $\rho _{ab}$ was measured by a
standard four-probe configuration within MPMS using an LR-700 resistance
bridge (frequency = 16 Hz, current = 1 - 3 mA).

\begin{figure}[tbp]
\centerline{\includegraphics[scale=0.35]{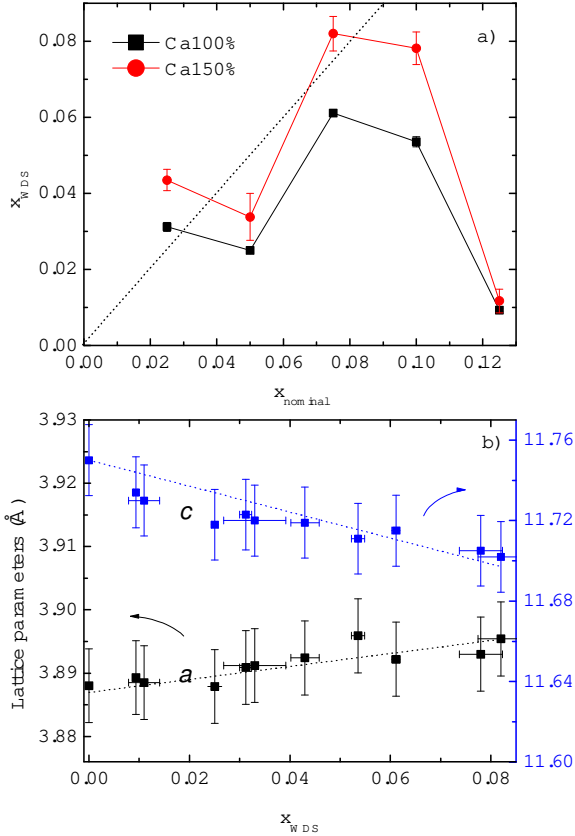}} \vspace*{-0.3cm}
\caption{a) Actual Co concentration as a function of nominal one. Black
squares represent the series with stoichiometric starting composition; red
circles represent the series with 50\% excess Ca in starting composition.
Dotted line indicates the ideal slope equal to 1. b) lattice parameters vs.
Co doping. Lines are guide to the eye.}
\end{figure}

Figure 1(a) shows the actual concentration of Co, $x_{WDS}$, as a function
of the nominal $x_{nominal}$, of two series of crystals: using
stoichiometric, nominal composition, polycrystalline feedstock, and using
polycrystalline feedstock that had 50\% excess Ca. The compositional spread
of the ten measured spots is taken as the error bar. In contrast to the
results in Ref. 16, $x_{WDS}$ deviates from linear dependence on $%
x_{nominal} $. After the elimination of the competing CaFe$_{4}$As$_{3}$
phase with 50\% excess Ca, $x_{WDS}$ generally increases and has a larger
compositional spread, but the curve follows the same trend as the
stoichiometric one and the significant non-monotonicity is still present. It
is noteworthy that for $x_{nominal}$ greater that 0.10, the corresponding $%
x_{WDS}$ decreases dramatically. This behavior suggests difficulties
associated with solubility and once again highlights the need to perform WDS
measurements on the grown samples. When the lattice parameters (from both
series) are plotted as a function of $x_{WDS}$ (Fig. 1(b)), there is a clear
linear dependence of $a$ and $c$ parameters on Co-substitution level. The
lattice parameters refined by Rietica are shown in Fig. 1(b). Lattice
parameter \textit{a} increases by 0.1\% whereas \textit{c} decreases by
0.4\% for $x_{WDS}=0.082$, similar to the trend in Sr(Fe$_{1-x}$Co$_{x}$)$%
_{2}$As$_{2}$. This is also in agreement with the results of Ref. 16, where 
\textit{c }linearly decreases, at $x_{EDX}=0.09$ $\Delta c/c=0.5\%$,
although we do not obtain samples with Co doping higher than $0.08$. There
are two pairs of concentrations very close to each other by coincidence,
i.e. $x_{WDS}=0.009,0.011$ and $0.031,0.033$. Only $x_{WDS}=0.009$ and $%
0.031 $ samples were characterized in the following study.

In-plane magnetic susceptibility of Ca(Fe$_{1-x}$Co$_{x}$)$_{2}$As$_{2}$ is
shown in Fig. 2(a) for magnetic field of 10 kOe. The structural/magnetic
transition of the parent CaFe$_{2}$As$_{2}$ at 170 K is suppressed
progressively by Co doping consistent with Ref. 16, until it is completely
suppressed at $x=0.054$. Superconductivity is first detected for $x=0.031$
and the superconducting transition temperature, $T_{c}$, decreases with
further substitution. Figure 2(b) shows the zero-field-cooled (ZFC) magnetic
susceptibility curves in an applied field of 100 Oe. The small dip at 3.5 K
for some curves is due to small Sn flux droplets on the crystal surface. The
superconducting shielding fraction, with the contribution from Sn
subtracted, varies with doping levels. The highest $T_{c}$ occurs at $%
x=0.043 $ whereas the largest volume fraction is reached at $x=0.054$ with
slightly lower $T_{c}$ (Fig. 2(b) inset). This is similar to what is
observed in Ref. 16.

\begin{figure}[tbp]
\centerline{\includegraphics[scale=0.35]{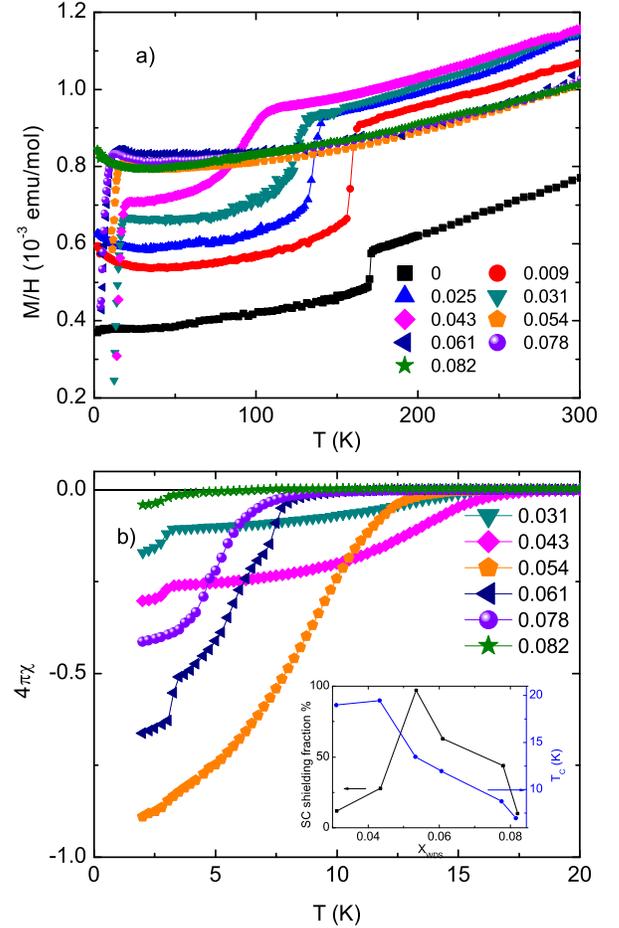}} \vspace*{-0.3cm}
\caption{a) In-plane magnetic susceptibility measured in a magnetic field of
10 kOe as a function of temperature. b) Zero-field-cooled magnetic
susceptibility in 50 Oe. Inset shows the variation of superconducting
shielding fraction and $T_{c}$.}
\end{figure}

Electrical resistance data, normalized to their room temperature values, are
shown in Fig. 3(a) for the Sn flux grown crystals. The anomaly at 170 K for
the pure CaFe$_{2}$As$_{2}$ is suppressed with Co doping, remains sharp
until $x_{WDS}=0.025$ and becomes broad for $x_{WDS}=0.031$ and $0.043$. (It
should be noted that this broadening coincides with the sudden onset of
superconductivity.) Figure 3(b) shows the low temperature normalized
resistance. The small jump at 3.5 K is due to remanent Sn flux on the
crystal. Although $x_{WDS}=0.031$ shows partial magnetic shielding and its
resistance starts to drop at about the same onset temperature, zero
resistance is not reached. Complete superconducting transition is observed
for $x_{WDS}\geq 0.043$ and $T_{c}$ gradually decreases with doping in good
agreement with the magnetic susceptibility measurements. Nanoscale
inhomogeneity and strain due to Co doping may result in wide transitions.
But considering the variation and small number of the superconducting volume
fraction, SC may not be bulk for many of the Ca(Fe$_{1-x}$Co$_{x}$)$_{2}$As$%
_{2}$ samples. Other experimental techniques, e.g. magneto-optical imaging,
specific heat, or STM spectroscopy will be required to further clarify the
nature/homogeneity of the low temperature state.

\begin{figure}[tbp]
\centerline{\includegraphics[scale=0.35]{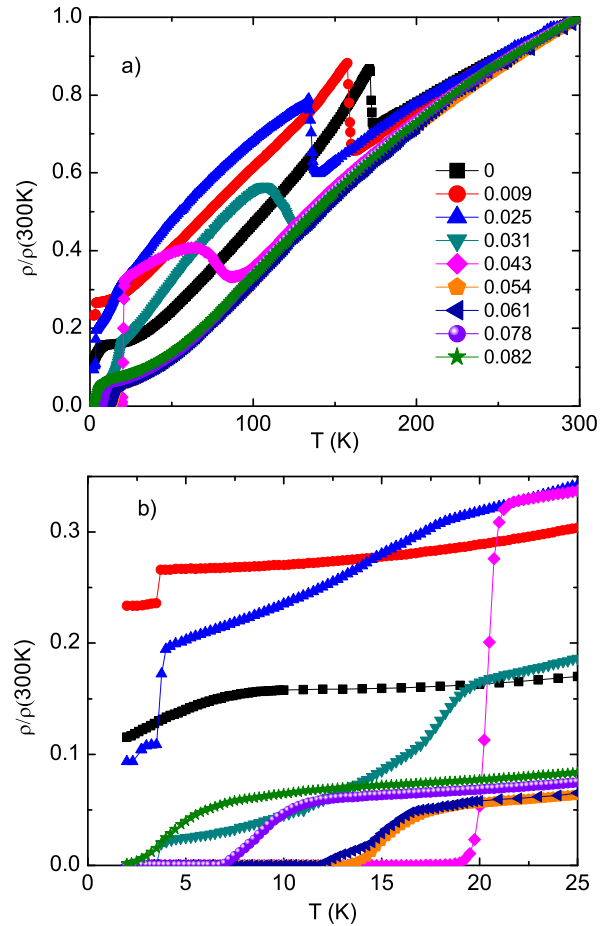}} \vspace*{-0.3cm}
\caption{a) Temperature dependence of the normalized resistivity of Ca(Fe$%
_{1-x}$Co$_{x}$)$_{2}$As$_{2}$. b) Expanded view of normalized resistivity
at low temperatures. }
\end{figure}

\begin{figure}[tbp]
\centerline{\includegraphics[scale=0.35]{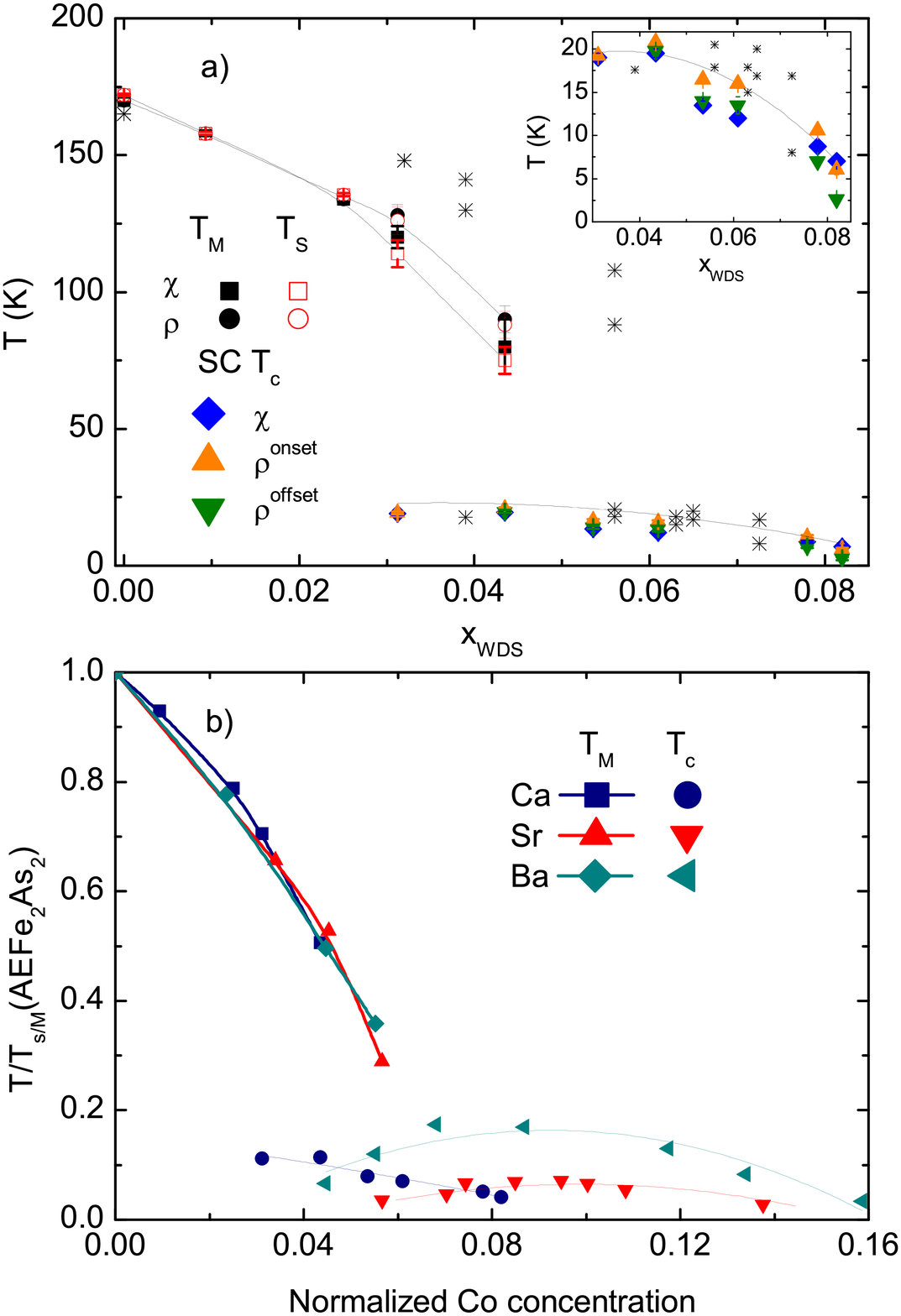}} \vspace*{-0.3cm}
\caption{a) $T-x$ phase diagram of Ca(Fe$_{1-x}$Co$_{x}$)$_{2}$As$_{2}$.
Solid lines are guides to the eye. Inset shows the superconducting region.
Black asterisks are the data from Ref. 16 inferred from resistance. b)
Comparison of the Ca (this work), Sr\protect\cite{Rongwei}, Ba\protect\cite%
{Ni} phase diagrams of Co doping. $T_{c}s$ are inferred from magnetic
susceptibility measurements. Normalization: $T$ axis is normalized by the $%
T_{S/M}$ of the respective parent compound; (Sr) $x$ axis multiplied by
0.81; (Ba) $x$ axis multiplied by 1.18.}
\end{figure}

Based on the magnetic and transport measurements, $T-x_{WDS}$ phase diagram
for Ca(Fe$_{1-x}$Co$_{x}$)$_{2}$As$_{2}$ is mapped out in Fig. 4(a). The
structural and magnetic transitions are inferred from $d\chi /dT$ and $%
d(\rho /\rho _{300K})/dT$ using the same criteria in Ref. 16.
Superconducting transition temperature $T_{c}$ is inferred from the first
deviation from the normal magnetic susceptibility of the ZFC curve.
Resistive onset and offset of $T_{c}$ values are inferred from the
intersects of the steepest slope with the normal state and zero resistance
respectively. The simultaneous structural and magnetic transition of the
pure CaFe$_{2}$As$_{2}$ is monotonically suppressed by Co doping, but as
seen from $\chi $ and $\rho $, the transition remains sharp for low dopings $%
x_{WDS}\leq 0.025$, no discernible splitting of both transitions can be
observed. For $x_{WDS}=0.031$ and $\ 0.043$, transition broadens and it is
possible to infer an upper structural transition and a lower magnetic
transition, similar to Ba(Fe$_{1-x}$Co$_{x}$)$_{2}$As$_{2}$.\cite{Ni} To
compare the reported phase diagram of Ca(Fe$_{1-x}$Co$_{x}$)$_{2}$As$_{2}$
in Ref. 16 with ours, we plot the data points (black asterisks) inferred
from resistance measurements of Ref. 16 in Fig. 4(a). As can be seen, though
our phase diagram shows a faster suppression of the magnetic and structural
transitions by Co doping, SC occurs roughly with the same $T_{c}$ in similar
region. It should be noted that the actual Co concentration reaches up to
0.15 in Ref. 16, but only onset of resistive or no superconducting
transition is observed above $x=0.09$, consistent with our observations. We
might imagine that an overestimate (underestimate) of the Co concentration
in Ref. 16 (our work) will shift the phase diagram.

Different from the superconducting dome in Sr(Fe$_{1-x}$Co$_{x}$)$_{2}$As$%
_{2}$\cite{Rongwei} and Ba(Fe$_{1-x}$Co$_{x}$)$_{2}$As$_{2}$\cite{Ni}, the
onset of SC in Ca(Fe$_{1-x}$Co$_{x}$)$_{2}$As$_{2}$ appears abruptly at high
temperature and gradually decreases with Co substitution. In order to
compare all three cases of Co doping, the magnetic transition boundaries of
Sr and Ba are collapsed on to that of Ca, namely the transition temperatures
are normalized by that of the pure parent AEFe$_{2}$As$_{2}$ and the Co
concentrations of Sr and Ba are scaled so as to get to a single manifold in
Fig. 4(b). Whereas both the Ba and Sr series manifest a maximum $T_{c}$
value close to the Co substitution level that drives the magnetic/structural
phase transitions to zero, the Ca series manifests maximum $T_{c}$ values
deep in the ordered region and has SC disappearing near the substitution
levels needed to suppress the antiferromagnetic/structural phase transition.
For the Co substituted Ca122 series the sudden onset of SC may instead be
correlated with the splitting of the structural/magnetic phase transition
that takes place for $0.025<x<0.031$. The reason for this difference is
currently not well understood, but may be related to the extreme pressure
and strain sensitivity of CaFe$_{2}$As$_{2}$ as a host material. Unlike Co
substituted BaFe$_{2}$As$_{2}$ or SrFe$_{2}$As$_{2}$, it is possible that
the changes in lattice parameter seen in Co substituted CaFe$_{2}$As$_{2}$
play a more important role in determining the phase diagram and represent an
additional term to the changes in band filling associated with Co
substitution.

In summary, Ca(Fe$_{1-x}$Co$_{x}$)$_{2}$As$_{2}$ ($0\leq x\leq 0.082$) have
been grown out of Sn flux. We report the details of single crystal growth
and their magnetic susceptibility and electrical transport properties. The
properties of single crystals are dependent on the growth procedure. The
phase diagram of Ca(Fe$_{1-x}$Co$_{x}$)$_{2}$As$_{2}$ shows a half dome like
superconducting region, different from those of Sr(Fe$_{1-x}$Co$_{x}$)$_{2}$%
As$_{2}$ and Ba(Fe$_{1-x}$Co$_{x}$)$_{2}$As$_{2}$. The electron doping as
well as chemical pressure probably are both responsible for determining the
phase boundary.

This work was carried out at the Iowa State University and supported by the
AFOSR-MURI grant \#FA9550-09-1-0603 (R. H. and P. C. C.). Part of this work
was performed at Ames Laboratory, US DOE, under contract \# DE-AC02-07CH
11358 (S. R., S. L. B. and P. C. C.). S. L. Bud'ko also acknowledges partial
support from the State of Iowa through Iowa State University.

*Present address: Center for Nanophysics \& Advanced Materials and
Department of Physics, University of Maryland, College Park MD 20742-4111,
USA.

\end{document}